\documentclass[12pt]{article}
\usepackage{graphicx}
\usepackage{amsmath}
\usepackage{amsfonts}
\usepackage{amsthm}
\usepackage{hyperref} 
\usepackage{bbm}
\usepackage{amsopn,amssymb,url,mathrsfs,a4wide}
\usepackage{tikz}
\usepackage{booktabs}
\usepackage[margin=2.25cm]{geometry}
\usepackage{parskip}
\usepackage{fancyvrb}
\usepackage{listings}
\usepackage[numbers]{natbib}
\usepackage{float}
\usepackage{verbatim}
\usepackage{caption}
\usepackage{subcaption}
\usepackage{titlesec}

\usepackage{natbib}
\setcitestyle{authoryear,open={(},close={)}}
\newcounter{ctr}

\newcounter{ctr1}

\newcounter{ctr2}

\newcounter{ctr3}

\newenvironment{theorem*}[1]{{\bf Theorem #1} \begin{itshape}}{\end{itshape}}

\newenvironment{corollary*}[1]{{\bf Corollary #1} \begin{itshape}}{\end{itshape}}

\newenvironment{proposition*}[1]{{\bf Proposition #1} \begin{itshape}}{\end{itshape}}

\newcommand{\ud}{\, {\rm d} \kern-.015em }

\newcommand{\modulus}[1]{\left| \kern.05em #1 \kern.05em \right|}
\newcommand{\norm}[1]{\left\| \kern.05em #1 \kern.05em \right\|}
\newcommand{\inner}[1]{\left\langle \kern.05em #1 \kern.05em \right\rangle }

\newcommand{\pick}[2]{\renewcommand{\arraystretch}{0.6}
\left( \kern-.4em \begin{array}{c} #1 \\ #2 \end{array} \kern-.4em \right) }

\newcommand{\var}[1]{\, {\rm var}\left( #1 \right) }

\newsavebox{\FVerbBox}
\usepackage{amsmath}
\usepackage{algorithm} 
\usepackage{setspace}
\usepackage{multirow}
\usepackage{algpseudocode}
\usepackage{cleveref}
\usepackage{listings}
\usepackage{xcolor}
\usepackage{centernot}
\newcommand*\sfref[1]{%
    Supplementary Figure \ref{#1}}
 
\DeclareMathOperator{\tr}{tr}

\providecommand{\keywords}[1]
{
  \small	
  \textbf{Keywords---} #1
}

\graphicspath{}
\newcommand{\commentOut}[1]{}
\title{Spatial deformation for non-stationary extremal dependence}
\author{J. Richards and J. L. Wadsworth}
\begin{document}
  \maketitle
\begin{abstract}
Modelling the extremal dependence structure of spatial data is considerably easier if that structure is stationary. However, for data observed over large or complicated domains, non-stationarity will often prevail. Current methods for modelling non-stationarity in extremal dependence rely on models that are either computationally difficult to fit or require prior knowledge of covariates. \cite{sampandgut} proposed a simple technique for handling non-stationarity in spatial dependence by smoothly mapping the sampling locations of the process from the original geographical space to a latent space where stationarity can be reasonably assumed. We present an extension of this method to a spatial extremes framework by considering least squares minimisation of pairwise theoretical and empirical extremal dependence measures. Along with some practical advice on applying these deformations, we provide a detailed simulation study in which we propose three spatial processes with varying degrees of non-stationarity in their extremal and central dependence structures. The methodology is applied to Australian summer temperature extremes and UK precipitation to illustrate its efficacy compared to a naive modelling approach. 
\end{abstract}
\hspace{10pt}
\keywords{non-stationary spatial dependence; extremal dependence; spatial deformation; max-stable processes}
  \section{Introduction}
  \label{intro}
Statistical methodology for spatial extremes can increasingly handle data sampled at more observation locations. If these observations are taken over large domains with complex features, then there is a strong chance that the data will exhibit spatial non-stationarity in both the marginal distributions and dependence structure. Marginal non-stationarity can often be dealt with by site-wise modelling and transformation. However, there are currently few methods to deal with non-stationarity in extremal dependence structures, and a typical approach is to falsely assume stationarity when fitting spatial extremes models. This may be appropriate when modelling data sampled over small and/or homogeneous regions in space, but as we will illustrate through the examples in Section \ref{simstudy}, this assumption is not realistic for many datasets with larger spatial domains.\par
Beyond site-wise transformation of margins, marginal non-stationarity can be handled by jointly modelling marginal parameters as functions of covariates. This can either be achieved parametrically \citep{Mannshardt_Shamseldin_2010, 10.2307/41714789, ribatet2013spatial, doi:10.1002/env.2560} or semiparametrically \citep{JONATHAN2014520, ROSS2017315,youngman, youngman2020flexible,doi:10.1002/env.2624} through the use of splines. Another widely applied approach is the use of Bayesian hierarchical models, in which the marginal parameters are assumed to come from some non-stationary latent process \citep{casson1999spatial,cooley2007bayesian,10.2307/20778450,Opitz2018INLAGE}.\par

Non-stationarity in the spatial dependence structure has been studied by \cite{huserNS} in the context of max-stable models, through incorporation of a non-stationary variogram. However, this approach requires knowledge of relevant covariates, and asymptotically dependent max-stable models for spatial extremes have been shown to be too inflexible for many spatial datasets \citep{invertref,Davison:189514,HUSER2017166,doi:10.1080/01621459.2017.1411813}. Another approach is to assume local stationarity for model fitting, see \cite{doi:10.1002/2017WR020717,doi:10.1080/01621459.2019.1647842}. This framework is well-suited to modelling processes with short-range dependence but is unlikely to fully capture dependence at large distances.   \cite{cooley2007bayesian} and \cite{10.2307/23069351} account for non-stationarity by transforming their spatial domain of interest to some new `climate space' in which observation locations with similar characteristics are grouped closer together. Again, this approach requires access to relevant covariates and a deeper understanding of the processes which are being modelled.
 \par 
 In this work we develop a computationally quick and simple method, which does not require prior knowledge of covariates and which can be applied before fitting any model suited to spatial extremes. Our method uses spatial deformation and is based on the work of \citet{sampandgut} and \citet{smith}, which has not been fully adapted for use in a spatial extremes framework. The deformation methodology may reveal physical features and/or covariates that can be incorporated into a spatial extremes model, removing the need for models with complex dependence structures.  \par
\cite{wadsworth2018spatial} applied the deformation method of \cite{smith} before fitting a conditional spatial extremes model to the same Australian summer temperatures data that we explore in Section \ref{secheat}. However, because this method is not tailored to extremal dependence, it was neccesary to assume that patterns in non-stationarity were similar for both the extremal and non-extremal dependence structures. \citet{youngman2020flexible} and \cite{chevalier2017modeling} provide extensions of the \cite{sampandgut} methodology and fit models for spatial extremes using deformations: a Gaussian process using a censored pairwise likelihood and a max-stable model, respectively. Although these models may be reasonable for some processes, use of either puts restrictions on the types of dependence that the process can exhibit. We look to develop a method that makes no strong assumptions on the extremal dependence structure.  \par
  The remainder of this section provides an overview of existing methodology for spatial deformation and modelling of spatial extremes. Our developments of the spatial deformation methodology are detailed in Section \ref{methodsec}. We present a simulation study in Section \ref{simstudy}, which is usually absent from the literature on spatial deformations. This study is used to convey that our adaptations to the deformation methodology are necessary when considering extremal dependence and that our method can be used for different processes with a wide range of extremal dependence structures. Finally, we apply our method to temperature and precipitation datasets in Section \ref{casestud}, and conclude with a discussion in Section \ref{discuss}.
  \subsection{Non-stationary spatial processes}
  \label{nspsec}
  The spatial deformation approach for handling non-stationarity in spatial processes was first proposed by \cite{sampandgut} and \cite{guttorp199420}, with further developments by \cite{Meiring97developmentsin}, see \citet[][Ch. 9.5]{gelfand2010handbook}. The underlying principle of their approach is that a smooth non-linear transformation can be used to map the sampling locations of a process from a geographical plane, or G-plane, to some latent space, which they name a D-plane, or dispersion plane. Within the D-plane, the dependence structure of the process is assumed to be both stationary and isotropic, and the usual statistical inferences can be made using stationary geostatistical models. To obtain the D-plane, optimisation techniques are used to minimise some objective function which is associated with a stationary geostatistical model. Here \citet{sampandgut} use multi-dimensional scaling and a stationary spatial dispersion function, whereas further work proposed by \citet{smith} uses the likelihood for a stationary Gaussian process. Our approach is to change this objective function for one which is associated with a stationary spatial extremes model, such as the max-stable, or inverted max-stable, processes.\par
 We begin by assuming we have realisations $\mathbf{Z}=\{Z_1,\dots,Z_N\}$ from a spatial field observed at sampling locations $s_1,\dots,s_d$, and so we have $Z_k=\{Z_k(s_1),\dots,Z_k(s_d)\}$ for all $k=1,\dots,N$. We require some smooth mapping function from the G-plane to the D-plane, given by $f(s_i)=s^*_i$ for $i=1,\dots,d$, where $s_i=(x_i,y_i)$ and $s^*_i=(x^*_i,y^*_i)$ are the corresponding locations in the D-plane. Both \citet{sampandgut} and \citet{smith} propose the use of thin-plate splines to achieve this mapping. However, we note that under certain conditions on the correlation structure, analytical forms for $f(\cdot)$ do exist. \cite{10.2307/3215594} prove that this mapping is identifiable assuming differentiability of the stationary and isotropic correlation function used for fitting and \cite{PERRIN200023} derive analytical forms for $f(\cdot)$ under the same assumption, with extensions to anisotropic correlation structures. As these results are available only for correlation functions, and not for extremal dependence functions, we instead use the more flexible thin-plate spline approach.\par
 A thin-plate spline is a mapping function $f(\cdot)$, passing through a finite number of data points $f^{*}_i=f^*(x_i,y_i),(i=1,\dots,n)$, minimising the bending energy
\[
J(f)=\iint_{\mathbb{R}^2}\left\{\left(\frac{\partial^2f}{\partial x^2}\right)^2+2\left(\frac{\partial^2f}{\partial x\partial y}\right)^2+\left(\frac{\partial^2f}{\partial y^2}\right)^2\right\}\mathrm{d}x\mathrm{d}y.
\]
Here we have denoted $f^{*}$ the `true' function that we wish to estimate with the thin-plate spline, $f$, and $f^*_i$ are observations.
\citet{greensilver} give a solution to this problem in the form
\begin{equation}
\label{splinefunc}
f(x,y)= a + bx + cy + \sum^n_{i=1}\delta_ig_i(x,y),
\end{equation}
where
\begin{equation}
\label{con}
\sum_{i=1}^n\delta_i=\sum_{i=1}^n\delta_i x_i=\sum_{i=1}^n\delta_i y_i = 0,
\end{equation}
and $g_i(x,y)=h_i^2\log h_i$, 
with $h_i$ the Euclidean distance between $(x,y)$ and $(x_i,y_i)$.
This represents $f$ as the sum of linear terms and $n$ radial basis functions with centres at the observed data locations $(x_i,y_i)$ and the constraints are in place to ensure that the system of equations does not become overdetermined. An interpolating spline satisfies $f^{*}_i=f(x_i,y_i)$ for all $i=1,\dots, n$, whereas we desire a smoothing spline; this can be created by minimising 
\[
S(f)=\sum_{i=1}^n\{f^{*}_i-f(x_i,y_i)\}^2+\alpha J(f),
\]
for some smoothing parameter $\alpha > 0$. \citet{sampandgut} give a method for estimating $\alpha$ in the context of multidimensional scaling, but here we take the approach of \citet{smith}, who uses a restricted representation of \eqref{splinefunc} instead. A subset of $m$ radial basis functions is used and so we let $\delta_i = 0$ for all $i\notin \{i_1,\dots,i_m\}$. The choice of this subset is discussed in Section \ref{simstudy}.\par
The function in \eqref{splinefunc} maps $\mathbb{R}^2$ to $\mathbb{R}$, so the spline is applied twice with different parameter estimates to produce both components. \citet{smith} gives a parametrisation as 
\begin{align}
\label{splinefunc2}
f^{(1)}(x,y)=  b_1^2x + \rho b_1b_2y + \sum^m_{i=1}\delta^{(1)}_ig_i(x,y)\\
f^{(2)}(x,y)=  b_2^2y + \rho b_1 b_2 x + \sum^m_{i=1}\delta^{(2)}_ig_i(x,y),
\end{align}
where $b_1>0,\; b_2>0,\; \rho \in \mathbb{R}$ and each of the sequences $\delta^{(1)},\;\delta^{(2)}$ satisfy the constraint in \eqref{con}. The introduction of the parameters $b_1,b_2$ and $\rho$ is to ensure that the model is invariant under orthogonal rotations when $m= 0$. Overall,  this yields a spline with $2m-3$ free parameters whenever $m \geq 3$.\par
The resulting spline is then used to map the sampling locations $s_i$ to locations $s^*_i$ in a latent space. Parameters are estimated by minimising some objective function provided by a stationary model. As previously mentioned, \citet{sampandgut} use a stationary spatial dispersion model and multidimensional scaling, the details of which are not given here. Instead, we focus on the approach by \citet{smith}, who uses a stationary Gaussian likelihood. It is assumed that $(Z(s^{*}_1),\dots,Z(s^{*}_d))\sim N_d(\mu,\Omega)$, where $\mu$ and $\Omega$ are the mean vector and a stationary covariance matrix, respectively.  As we are only interested in measuring the dependence structure, it is assumed that the means and variances at each location are known. Analysis is then simplified to only considering the minimisation of the negative log likelihood given by
\begin{equation}
\label{loglikSGP}
-\log L(\Omega)=\frac{N}{2}\log|\Omega|+\frac{N-1}{2}\tr\left(\Omega^{-1}\hat{\Omega}\right),
\end{equation}
where $\Omega$ and $\hat{\Omega}$ are the theoretical, and sample, correlation matrices and $\tr(\cdot)$ and $|\cdot|$ are the trace and determinant operators, respectively. The entries of the theoretical correlation matrix are produced by using a stationary covariance function. \citet{smith} uses the Mat\'ern covariance function, and so
\begin{equation}
\label{matern}
\Omega_{ij}=\frac{1}{2^{\theta_2-1}\Gamma(\theta_2)}\left(\frac{2h^*_{ij}\sqrt{\theta_2}}{\theta_1}\right)^{\theta_2}K_{\theta_2}\left(\frac{2h^*_{ij}\sqrt{\theta_2}}{\theta_1}\right),
\end{equation}
where $\theta_1>0,\; \theta_2>0$ and $K_{\theta_2}(\cdot)$ is the modified Bessel function of the second kind of order $\theta_2$ and $h^*_{ij}=\|s^*_i-s^*_j\|$ is the Euclidean distance between locations $s^*_i$ and $s^*_j$ in the D-plane. It is noted that $\theta_1$ can be set to 1 as the spatial scaling of the locations is controlled by the spline. 
\subsection{Spatial extremes}
Before describing an extension of the spatial deformation methodology tailored to spatial extremes, we first provide a brief review of methods for modelling spatial extremes.  
\subsubsection{Max-stable and inverted max-stable processes}
Max-stable processes were introduced by \cite{haan1984} and developed further by \cite{smith1990max} and \cite{schlather2002models}, who suggested models that were first fitted by pairwise composite likelihood in \cite{doi:10.1198/jasa.2009.tm08577}. They are usually described by a spectral construction. Suppose $\{r_i;i \geq 1\}$ are points of a Poisson process on $(0,\infty)$ with unit intensity. Let $S \subseteq \mathbb{R}^2$ be a spatial index set, and $\{W_i(s);s \in S, i \geq 1\}$ be independent and identically distributed copies of a non-negative stochastic process satisfying $\mathbb{E}[W(s)]=1.$ Then
\begin{equation}
\label{MSPrep}
Z(s)=\max_{i \geq 1}\left\{ W_i(s)/r_i \right\}
\end{equation}
is a max-stable process with unit Fr\'echet margins. The $d$-dimensional joint distribution function for $Z$ is 
\begin{equation}
\label{brjoint}
\Pr\{Z(s_1)\leq z_1,\dots,Z(s_d) \leq z_d\}=\exp\{-V(z_1,\dots,z_d)\},
\end{equation}
where the exponent is
\begin{equation}
\label{exponentV}
V(z_1,\dots,z_d)=\mathbb{E}\left[\max\left\{\frac{W(s_1)}{z_1},\dots,\frac{W(s_d)}{z_d}\right\}\right].
\end{equation}
Careful specification of the stochastic process $W(s)$ leads to a limited selection of parametric models for the max-stable process. A particularly flexible model is the Brown-Resnick model \citep{10.2307/3213346,madoc31285}. This involves specifying $W(s)=\exp\{U(s)-\gamma^*(s,0)\}$ for $U(s)$ a centred Gaussian process with semivariogram $\gamma^*(\cdot,\cdot)$ and where $U(0)=0$ almost surely. This leads to a 2-dimensional joint distribution with exponent function
\begin{equation}
\label{exponent}
V(z_i,z_j)=\frac{1}{z_i}\Phi\left\{\frac{a}{2}-\frac{1}{a}\log\left(\frac{z_i}{z_j}\right)\right\}+\frac{1}{z_j}\Phi\left\{\frac{a}{2}-\frac{1}{a}\log\left(\frac{z_j}{z_i}\right)\right\},
\end{equation}
where $a = [2\gamma^*(s_i,s_j )]^{1/2}$ and $\Phi(\cdot)$ denotes the standard normal distribution function. Note that for a stationary and isotropic Brown-Resnick process, $\gamma^*(s_i,s_j)$ is dependent on $h_{ij}=\|s_i-s_j\|$ only. For clarity, we write $\gamma(h_{ij})$ when $Z$ is stationary and isotropic, and $\gamma^*(s_i,s_j)$, otherwise. Representations for \eqref{exponent} in higher dimensions exist (see \cite{brpro} or \cite{10.1093/biomet/ast042}), but due to their computational complexity, inference for max-stable processes is typically done pairwise, providing a reasonable balance between computation time and efficiency.\par
Max-stable processes are inherently asymptotically dependent, or perfectly independent. That is, $Z(s_i)$ and $Z(s_j)$ are asymptotically dependent, or perfectly independent, for all $s_i, s_j \in S$. Here we characterise asymptotic dependence using the upper tail index $\chi$ \citep{joe1997multivariate}. Assuming $Z(s_i)\sim F_i,Z(s_j)\sim F_j$, we have
\begin{equation}
\label{chidef}
\chi(s_i,s_j)=\lim_{q\rightarrow 1}\Pr\{F_i\{Z(s_i)\}>q|F_j\{Z(s_j)\}>q\},
\end{equation}
where the process is asymptotically independent at locations $s_i$ and $s_j$ if $\chi(s_i,s_j)=0$, and asymptotically dependent otherwise. Here we write $\chi(s_i,s_j)$ as $Z$ is not necessarily stationary; henceforth, we write $\chi(h_{ij})$ for $h_{ij}=\|s_i - s_j\|$ when it is assumed that $\chi$ is a function of distance only.  As this measure is theoretically non-zero at all spatial lags for any max-stable process exhibiting positive spatial association ie., $\chi(s_i,s_j) > 0$  for all $s_i,s_j \in S$, we require other modelling approaches to deal with processes that may exhibit asymptotic independence.\par
\citet{invertref} introduced the inverted max-stable process as that obtained by applying a monotonically decreasing marginal transformation to a max-stable process. For example, with $Z$ as defined in \eqref{MSPrep}, taking $Y(s)=1/Z(s)$ gives an inverted max-stable process with exponential margins and joint survival function
\begin{equation}
\label{ibrjoint}
\Pr\{Y(s_1)\geq y_1,\dots,Y(s_d) \geq y_d\}=\exp\{-V(1/y_1,\dots,1/y_d)\},
\end{equation}
where $V$ is as given in \eqref{exponentV}. Such a process is asymptotically independent with $\chi(s_i,s_j) = 0$ for all $s_i\neq s_j$, but can accommodate a variety of flexible extremal dependences structures exhibiting positive association. The dependence in asymptotically independent processes may be characterised by a pre-limiting version of \eqref{chidef}. Specifically, under an assumption of hidden regular variation \citep{ledford1996statistics,resnick2002hidden},
\begin{equation}
\label{etaeq}
\chi_{q}(s_i,s_j)=\Pr\{F_i\{Z(s_i)\}>q|F_j\{Z(s_j)\}>q\}=L(1-q)(1-q)^{1/\eta(s_i,s_j)-1},
\end{equation}
with $L(\cdot)$ slowly varying at 0 and $\eta(s_i,s_j) \in (0,1]$ the coefficient of tail dependence. For an inverted max-stable process, $\chi_q(s_i,s_j)=(1-q)^{V(1,1)-1}$.
\par
We fit both max-stable and inverted max-stable models after applying our deformation method for non-stationary spatial extremes. Note that although max-stable processes are typically taken to represent the limiting behaviour of maxima, in practice they, along with inverted max-stable processes, can be used for all extreme values through specification of a censored likelihood; see Section \ref{fitsec}. Inference on these models can then be used to determine the efficacy of our deformation method.
\subsubsection{Conditional extremes}
\label{ConExSec}
An alternative approach to modelling spatial extremes is to condition on the behaviour of the process when it is extreme at a single site. Here we give a brief overview of modelling the extremal behaviour of the process at two sites using this approach. For a full characterisation, see \cite{wadsworth2018spatial} or \cite{robwaves}. We suppress some of the notation used by \cite{wadsworth2018spatial} and \cite{robwaves} as we are only considering a discrete pairwise fit, that we will employ in Section \ref{casestud} as a diagnostic measure. For further details of the discrete approach, see \cite{hefftawn}. \cite{hugo} apply this same methodology to a dataset of Australian temperatures, which we revisit in Section \ref{secheat}.  \par

We begin by assuming that $\{X(s):s \in S \subset \mathbb{R}^2\}$ is a stationary and isotropic process with exponential-tailed marginals and denote $X(s_i)=X_i$. Conditioning on $X_i=x_i>u$ being large and considering $X_j, i\neq j$, \citet{hefftawn} assume that there exist normalising functions $a(x_i):\mathbb{R}\rightarrow \mathbb{R}, b(x_i):\mathbb{R}\rightarrow\mathbb{R}_{+}$, for which
\[
\lim_{x_i \rightarrow \infty}[\Pr(X_j \leq a(x_i)+b(x_i)z|X_{i}=x_{i})]=G(z),
\]
where $G$ is non-degenerate. Re-writing $Z=\{X_j-a(x_i)\}/b(x_i)$ as the standardised residual,
and making the assumption that the limit holds above some high threshold $u$, we have
\[
\Pr(Z\leq z|X_{i}=x_{i})=G(z),\;\;  x_i>u,
\]
where $X_i|X_i>u \sim \text{Exp}(1)$ is independent of $Z$. Inference on $G$ is often simplified by making the working assumption that $Z \sim N(\mu, \sigma^2)$ and using a specified parametric form for the normalising functions $a(\cdot),b(\cdot)$. For positively dependent data, we simplify the normalising functions to $a(x_i)=\alpha x_i$ for $\alpha \in[0,1]$ and $b(x_i)=x_i^{\beta}$ for $\beta \in [0,1)$. The bivariate form of the conditional model can thus be expressed
\[
X_j|\left(X_i=x_i\right)=\alpha x_i+x_i^{\beta}Z,\;\;\;\;x_i>u.
\]
The conditional model holds some useful advantages over joint modelling using max-stable, or inverted max-stable, processes. For one, it is able to handle both asymptotically dependent, or asymptotically independent, data. Parameter estimates for $\alpha$ and $\beta$ can indicate the nature of the dependence between $X_j$ and $X_i$. For example, asymptotic dependence between $X_j$ and $X_i$ is implied by estimates  $\alpha=1, \beta=0$. Within the class of asymptotically independent variables, $\alpha<1, \beta>0$, with $\alpha=\beta=0$ giving near extremal independence.\par The spatial extensions of this model \citep{wadsworth2018spatial,robwaves} specify  $\alpha$ and $\beta$ as functions of distance between sites, when the underlying process is stationary and isotropic. As such, we can use these parameter estimates as diagnostics, to determine whether our deformation method has created a process that has a more stationary extremal dependence structure. We are motivated to use these estimates as our deformation method does not use a conditional extremes approach for fitting.
  \section{Spatial deformation for extremes}
  \label{methodsec}
  In this section, we discuss our adaptations of the deformation methodology for application in a spatial extremes framework. We begin in Section \ref{secobjec} by proposing a new objective function to that of \eqref{loglikSGP}. Instead, we consider minimising the difference between theoretical and empirical $\chi$ measures, where the former are produced through specification of a stationary max-stable dependence structure for the process in the D-plane. This does not in fact mean that this method will not work for asymptotically independent data; on the contrary, in Sections \ref{secmodelspec} and \ref{secchoice} we show that the model choice for $\chi(\cdot)$ is somewhat arbitrary and a single, simple parametric form works well for both classes of extremal dependence.  Section \ref{secpract} follows with some practical advice for choosing the anchor points used in estimating the thin-plane spline and we conclude with details of model fitting and selection using censored pairwise likelihoods in Section \ref{fitsec}. To assess the efficacy of the deformations we produce, we fit full max-stable, and inverted max-stable, dependence models.
  \subsection{Objective function}
  \label{secobjec}
  To adapt the methodology of \citet{sampandgut} and \citet{smith} to better suit a spatial extremes framework, we change the objective function given in \eqref{loglikSGP} to the Frobenius norm of the difference between theoretical and empirical pairwise dependence matrices $\mathrm{X}:=[\chi(h^{*}_{ij})]$ and $\hat{\mathrm{X}}:=[\hat{\chi}(h^{*}_{ij})]$. That is, we estimate the parameters of the thin plate spline through computing
  \begin{equation}
  \label{frob}
  \min \|\mathrm{X}-\hat{\mathrm{X}}\|_F = \min\sqrt{\sum_{i=1}^d\sum^d_{j=1}\left\{\chi(h^{*}_{ij})-\hat{\chi}(h^{*}_{ij})\right\}^2},
  \end{equation}
where $\chi(h^{*}_{ij})$, defined in \eqref{chidef},
is the upper tail index calculated between the process at locations $s^*_i$ and $s^*_j$ in the D-plane and $\hat{\chi}(h^{*}_{ij})$ is its empirical estimate. Recall that we assume stationarity in the D-plane, and so write $\chi(h^*_{ij})$, rather than $\chi(s^*_i,s^*_j)$. In practice, this measure cannot be estimated in the limit as $q \rightarrow 1$. As such, we estimate $\hat{\chi}(h^{*}_{ij})$ 
by fixing some high threshold $q<1$ and calculating
\begin{equation}
\label{chiestim}
\hat{\chi}_{q}(h^{*}_{ij})=\Pr\{\hat{F}_i\{Z(s^{*}_i)\}>q|\hat{F}_j\{Z(s^{*}_j)\}>q\}=\Pr\{\hat{F}_i\{Z(s_i)\}>q|\hat{F}_j\{Z(s_j)\}>q\},
\end{equation}
where $\hat{F}_k(\cdot)$ is the empirical distribution of observations $Z(s^{*}_k)=Z(s_k)$. Under asymptotic dependence, we assume that $\chi_{q}(h^{*}_{ij}) \equiv \chi(h^{*}_{ij})$ for large enough $q$. Under asymptotic independence, although $\chi_q(h^{*}) \rightarrow 0$ as $q \rightarrow 1$, we typically have $\chi_q(h^{*})>0$ for $q<1$ and spatial structure in this measure that makes it informative about non-stationarity. \par
We now focus on a choice of function $\chi(h^{*})$, which we only require to be monotonically decreasing from $1$ to $0$. This leaves several options, including specific parametric forms for $\chi(h^{*})$ and $\chi_q(h^{*})$ from max-stable, and inverted max-stable, processes.
We remark that while we have used $\chi$ to measure extremal dependence, other extremal dependence measures exist, and can also be used in this framework. For example, the coefficient of tail dependence, $\eta(h^{*}_{ij})$, from \eqref{etaeq} can also be used to characterise the strength of asymptotic independence in extremes. This can be estimated separately from $\chi(h^{*}_{ij})$, however, we found that due to the high variance of the estimator for $\eta(h^*_{ij})$, it was often outperformed by using $\chi(h^{*}_{ij})$. 
\subsection{Asymptotic dependence versus asymptotic independence}
\label{secmodelspec}
\begin{figure}[h]
 
  \centering
  
 \includegraphics[width=\linewidth]{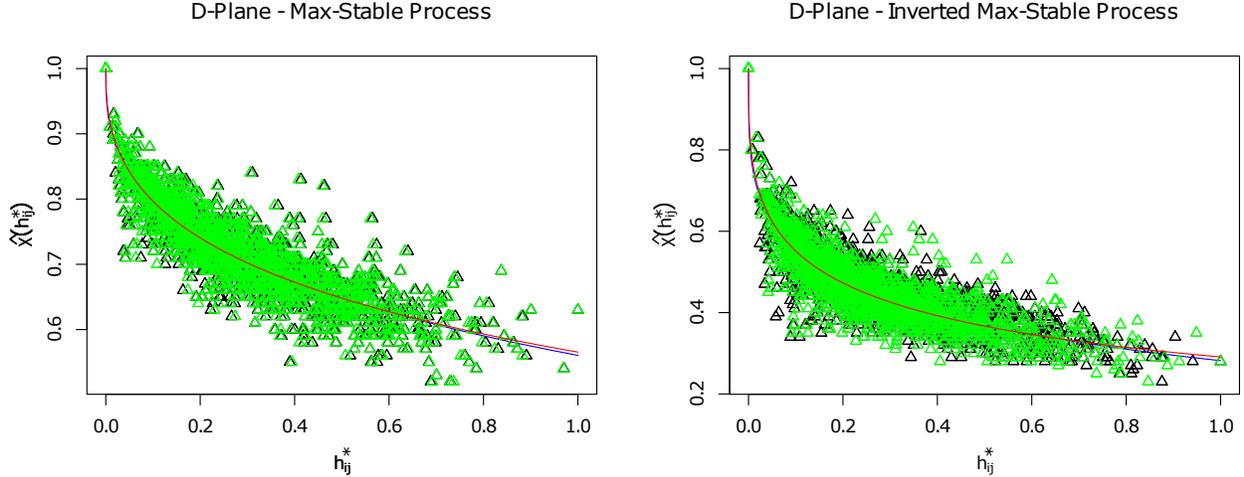} 
 
  \caption{Comparison of deformations created using both parametric forms $\chi^{BR}$ and $\chi_q^{IBR}$ for $\chi(\cdot)$ for both max-stable data (left) and inverted max-stable data (right). Plots show empirical  $\chi(h^{*}_{ij})$ estimates against distance, where the black triangles correspond to those created using $\chi(\cdot)$ given by \eqref{chiBR} and green triangles for those created using \eqref{chiIBR}. The blue and red lines give the fitted function from \eqref{chiBR} and \eqref{chiIBR}, respectively. Distances are normalised so that the maximum distance is consistent between deformations.}
   \label{chiBRvschiIBR}
  \end{figure}

As a parametric model for $\chi(h^{*})$ we take the form implied by the stationary Brown-Resnick process,
\begin{equation}
\label{chiBR}
\chi(h^{*}_{ij})=2-\theta(h^{*}_{ij})=2-2\Phi\left\{\frac{[2\gamma( h^*_{ij})]^{1/2}}{2}\right\},
\end{equation}
where $\theta(\cdot)$ is the extremal coefficient function \citep{schlather2003dependence} and $\theta(h^{*}_{ij})=V(1,1)$, with $V(\cdot,\cdot)$ defined in \eqref{exponent}. The semivariogram $\gamma(h^*_{ij})$ controls the dependence of the max-stable field and a typical choice for the semivariogram would be 
\begin{equation}
\label{vario}
\gamma( h^*_{ij})=(h^*_{ij}/\lambda)^\kappa,
\end{equation}
where $\lambda >0$ is a scaling parameter and $\kappa \in (0,2]$ is a smoothing parameter. Note that setting $\kappa = 2$ yields the Smith process \citep{smith1990max}, a special case of the Brown-Resnick process. As previously mentioned when discussing the \citet{smith} methodology for spatial deformation, we can set the scaling parameter $\lambda$ to 1, as the spatial scaling of locations is controlled by the deformation itself.
Note that the motivation for using the Brown-Resnick process as a parametric model is that $\chi(h^{*}) \rightarrow 0$ as $h^{*}\rightarrow \infty$, unlike other popular parametric models.
For a stationary inverted Brown-Resnick process, we have 
\begin{equation}
\label{chiIBR}
\chi_q(h^{*}_{ij})=(1-q)^{\theta(h^{*}_{ij})-1}.
\end{equation}
 We denote the dependence measures in \eqref{chiBR} and \eqref{chiIBR} as $\chi^{BR}$ and $\chi^{IBR}_{q}$, respectively. Note that although these two measures have different parametric forms, and are applicable to different dependence structures, they often approximate each other very closely when used within a deformation framework; this is illustrated in Figure \ref{chiBRvschiIBR}. Here we create deformations for a simulated dataset as described in Section \ref{NSMSPsec} using both $\chi^{BR}$ and $\chi^{IBR}_{q}$. The plots show that both methods give very similar deformations when considering the non-stationarity in the $\chi(h^{*}_{ij})$ estimates. This seems to be the case for both asymptotically dependent and asymptotically independent data.  Hence, for the sake of simplicity we only use $\chi^{BR}$ to create deformations in the case studies in Section \ref{casestud}, as it appears to be flexible enough to capture non-stationarity in both classes of extremal dependence.\par
  \subsection{Choice of parametric model for \texorpdfstring{$\chi(h^{*})$}{x(h^{*})}}
  \label{secchoice}
We have also found that the function $\chi(h^{*})$ from a Brown-Resnick process is sufficiently flexible to create suitable deformations for a variety of different extremal dependence structures. This is for similar reasons to above; different functions $\chi(h^{*})$ which decrease to zero as $ h^{*}\rightarrow \infty$ can approximate each other well. To illustrate this, we also considered the Gaussian-Gaussian process \citep{invertref}, which encompasses different dependence structures to the Brown-Resnick process, but for which $\chi(h^{*}) \rightarrow 0$ as $ h^{*} \rightarrow \infty$. Its theoretical form is
\[
\chi^{GG}(h^{*})=1-\frac{1}{2}\int_{\mathbb{R}^2}\{\phi(u)^2-2\rho(h^{*})\phi(u)\phi(u-\tilde{h})+\phi(u-\tilde{h})^2\}^{1/2}\mathrm{d}u,
\]
where $\rho(h^{*})$ is a stationary correlation function and $\tilde{h}=(h^*, h^*)^T$ and $\phi(\cdot)$ is the bivariate Gaussian density function with mean $0$ and covariance matrix $\Sigma=\text{diag}(\sigma^2,\sigma^2)$. Note that using a Mat\'ern correlation function given in \eqref{matern} with parameters $\theta_1>0$ and $\theta_2>0$, this function has one extra parameter than $\chi^{BR}(h^{*})$, namely $\sigma>0$.\par
We chose not to use this parametric form for $\chi(h^{*})$, due to the high computational cost required to compute the double integral for each pair of locations. However, we have found that the deformation method described in Section \ref{methodsec} appears fairly robust to the choice of $\chi(h^{*})$. As \sfref{GG} in the Supplementary Material shows, the much simpler $\chi^{BR}(h^{*})$ can approximate the more complex $\chi^{GG}(h^{*})$ very closely for much of $h^{*} \in \mathbb{R}_+$.
\subsection{Practical aspects for creating deformations}
\label{secpract}
We now comment on practical aspects of creating the deformations, including choosing a subset of radial basis functions for the thin-plate spline and reducing the chances of producing a non-bijective transformation.\par
We found that there is no simple robust method for picking the number $m$, or configuration, of the anchor points used in the deformation splines given in \eqref{splinefunc2}. As detailed in \citet{sampandgut}, there is a trade-off in picking $m$. Larger values provide ``better'' deformations, in the sense that the objective function to be calculated is lower and the deformations seem to capture more of the non-stationarity in the process. However, this comes at the price of computational cost, the risk of over-fitting and the phenomenon in which the D-plane folds on to itself. This provides a non-bijective transformation, which is physically unrealistic. \cite{doi:10.1198/1061860043100} detail an approach to ensure that the deformation is always bijective through use of a simulated annealing algorithm, with later extensions by \cite{youngman2020flexible}. These approaches add further constraints into the modelling procedure, which we have chosen to avoid. Instead we use a more heuristic approach for avoiding non-bijectivity.\par
We begin by randomly sampling $m_0$ initial anchor points with index set given by $I_0=\{i_1,\dots,i_{m_0}\}$. There is no single best way to choose $I_0$; however, we found that ensuring that the anchor points are spread out over the spatial domain helped to create better deformations. Performing a deformation with $I_0$ yields parameter estimates $\hat{\boldsymbol{\psi}}_0=(\hat{b}_1,\hat{b_2},\hat{\rho},\hat{\kappa},\hat{\delta}^{(1)}_4,\hat{\delta}^{(2)}_4,\dots,\hat{\delta}^{(1)}_{m_0},\hat{\delta}^{(2)}_{m_0})$. Recall that we have parameters $\delta^{(1)}_i,\;\delta^{(2)}_i$ indexed by $i\geq 4$ as those indexed by $i=1,2,3$ are uniquely determined by the constraints given in \eqref{con}. If the deformation for $I_0$ is bijective, we create a new set of indices $I_1=\{I_0,i_{m_0+1}\}$, where $i_{m_0+1}$ is sampled from the remaining indices. A deformation is then created using $I_1$, but with initial parameters in the optimisation program given by $\hat{\boldsymbol{\psi}}_1=\{\hat{\boldsymbol{\psi}}_0,\delta^{(1)}_{m_0+1}=0,\delta^{(2)}_{m_0+1}=0\}$. This ensures that the initial input into the optimisation program creates a deformation that is already bijective. We then continue in this fashion until we have created a deformation using $m^{*}$ anchor points. Bijectivity is checked by eye.\par
Using this approach reduces the chances of the D-plane folding as $m$ increases and provides a deformation with $m^*$ anchor points. Here we set $m^*$ as approximately a quarter of the sampling locations as we have not found a clear way to optimize this aspect. Typically this approach can be used for a number of initial index sets. However, in the interest of reducing computational cost, the simulation studies in Section \ref{simstudy} are conducted using the same initial index set for each deformation method. We also ensure that the new index sampled at each iteration is consistent across different samples, processes and deformation methods. 
\subsection{Model fitting and selection}
\label{fitsec}
To determine whether the deformation has created a process that is more stationary in the extremal dependence structure, and to compare between deformation methods, we look to fitting max-stable and inverted max-stable models to the data using the sampling locations in both the G-plane and the D-plane.
In Section \ref{intro}, the computational complexities of the max-stable and inverted max-stable models were discussed. To accommodate for this, we take a pairwise composite likelihood approach and assume independence between pairs \citep{doi:10.1198/jasa.2009.tm08577}. The joint distribution for a Brown-Resnick process is given in \eqref{brjoint} and the joint survival function for an inverted Brown-Resnick process is given in \eqref{ibrjoint}.
Note that the former is on standard Fr\'echet margins, whereas the latter is on standard exponential. To compare between the asymptotically dependent and asymptotically independent structures provided by the two models, we calculate all likelihoods on exponential margins, by first using a site-wise empirical transformation.\par
Given realisations $\{z_1,\dots,z_N\}$ from a spatial field, observed at sampling locations $s_1,\dots,s_d$, the censored composite likelihood is
\begin{equation}
\label{lik}
L_{CL}(\lambda,\kappa)=\prod^N_{i=1}L_{CL}(\lambda,\kappa;z_i)=\prod^N_{i=1}\prod_{k=2}^d\prod_{l<k}g_u(z_i(s_k),z_i(s_l);\lambda, \kappa),
\end{equation}
where
\begin{equation}
\label{compLikeq}
g_u(z_i(s_k),z_i(s_l);\lambda,\kappa)=\begin{cases}
f(z_i(s_k),z_i(s_l);\lambda,\kappa)\;\;&\text{if}\;\;\min(z_i(s_k),z_i(s_l))> u,\\
\frac{\partial}{\partial z_i(s_k)}F(z_i(s_k),u;\lambda,\kappa)\;\;&\text{if}\;\;z_i(s_k) > u, z_i(s_l) \leq u,\\
\frac{\partial}{\partial z_i(s_l)}F(u,z_i(s_l);\lambda,\kappa)\;\;&\text{if}\;\;z_i(s_k) \leq u, z_i(s_l) > u,\\
F(u,u;\lambda,\kappa)\;\;&\text{if}\;\;\max(z_i(s_k),z_i(s_l))\leq u,
\end{cases}
\end{equation}
with $u$ a high threshold and $F(\cdot)$ and $f(\cdot)$ the bivariate joint distribution and density functions for the model. Note that although we set $\lambda = 1$ when producing the deformation, here we treat it as a free parameter. Although the likelihoods give a good indication of the performance of the deformation methods, we use the Composite Likelihood version of the Akaike Information Criterion (CLAIC) for model selection. As given in \citet{10.2307/24309261}, the CLAIC is
\begin{equation}
\label{CLAICeq}
-2\{\log L(\hat{\lambda},\hat{\kappa})-\tr(J(\hat{\lambda},\hat{\kappa})H^{-1}(\hat{\lambda},\hat{\kappa}))\},
\end{equation}
where $(\hat{\lambda},\hat{\kappa})$ are the maximum likelihood estimates from \eqref{lik}, $H(\cdot)$ is the Hessian matrix and $J(\cdot)$ is the variance of the score function, i.e.
\[
J(\hat{\lambda},\hat{\kappa})=\var{\nabla\log L_{CL}(\hat{\lambda},\hat{\kappa})}=\var{\sum^N_{i=1}\nabla\log L_{CL}(\hat{\lambda},\hat{\kappa};z_i)}.
\]
In practice, we estimate $J(\cdot)$ by using numerical methods to find $\Delta_i=\nabla\log L_{CL}(\hat{\lambda},\hat{\kappa};z_i)$,
and then estimating the variance of the score function by setting a block of length $b<N$ and computing
\begin{equation}
\label{scoreest}
\hat{J}(\hat{\lambda},\hat{\kappa})=\frac{N}{b} \times \var{\sum^b_{i=1}\Delta_i,\dots,\sum^{N}_{i=N-b+1}\Delta_i} .
\end{equation}
The block sizes are chosen such that each block of data is more reasonably assumed approximately independent. This is usually specific to the data and will be given alongside any results.
  \section{Simulation study}
  \label{simstudy}

  We conduct three simulation studies to illustrate the efficacy of the deformation framework for modelling extremal dependence of non-stationary spatial processes. These studies are designed to highlight the following:
  \begin{itemize}
\item When fitting a stationary model to the extremal dependence of non-stationary spatial data, using a deformation method will improve the fit when compared to using the original sampling locations in the G-plane;
\item The deformation methodology described in Section \ref{secobjec} is more effective than the original \citet{smith} method when modelling non-stationary extremal dependence, as the latter is tailored towards modelling dependence in the body of the data rather than the extremes;
\item It is often necessary to use a deformation method that is tailored explicitly to extremal dependence, rather than dependence throughout the body; especially for processes that exhibit different degrees of non-stationarity throughout their extremal and central dependence structures.
\end{itemize}

In order to illustrate these points, we consider five different processes. These processes are chosen as they each exhibit different behaviour in their respective extremal dependence structures. In Section \ref{NSMSPsec}, we consider two processes: a non-stationary Brown-Resnick process and a non-stationary inverted Brown-Resnick process. In Section \ref{maxmixsec}, we consider two more processes which are both mixtures of stationary and non-stationary processes. We term these max-mixture process and one exhibits asymptotic dependence whilst the other exhibits asymptotic independence. A final process is considered in Section \ref{gaussmixproc}, which is an asymptotically independent Gaussian mixture process.
\par

For each setting, we begin with a sample of 1000 realisations of a spatial process. For this sample, we create four separate deformations using the procedure set out in Section \ref{secpract}. The first two deformations are created using the approach detailed in Section \ref{secobjec}; with $\chi^{BR}$ from \eqref{chiBR} and $\chi_{q}^{IBR}$ from \eqref{chiIBR} as the dependence measures used in the objective function in \eqref{frob}. The latter two are correlation-based deformation methods: one of these is the original \citet{smith} methodology, while the other method replaces $\chi(h^{*}_{ij})$ in \eqref{frob} with pairwise correlation $\rho(h^{*}_{ij})$ as the dependence measure, and replaces the theoretical $\chi(h^*)$ function with the stationary Mat\'ern correlation function detailed in \eqref{matern}. Note that in both of the latter two methods, correlation is estimated on a Gaussian marginal scale, and for the former two methods, we set $q = 0.9$ in \eqref{chiestim} and \eqref{chiIBR}.
 \par 
  
As detailed in Section \ref{fitsec}, we evaluate the efficacy of each of the four deformations by fitting a model to the extremal dependence of the sample. We fit the same dependence model five times: once using the sampling locations in the original G-plane and then once for each of the respective D-plane sampling locations given from the four deformations. For each fitted model, we calculate the CLAIC given in \eqref{CLAICeq}. Ordering of the CLAIC allows us to determine which deformation method (if any) was the most effective in accounting for the non-stationarity in that  sample. As the underlying process from which the sample is drawn is known, we fit a stationary extremal dependence model of an appropriate class. That is, for processes that are asymptotically dependent, we fit a stationary Brown-Resnick model, and for processes that are asymptotically independent, we fit a stationary inverted Brown-Resnick model. 
\par

This procedure is repeated for 50 different samples of a single process. In this simulation study, each deformation for each sample is created using the same anchor points. For each sample, we determine which deformation method was the most effective and the proportion of times this occurred over all samples is reported, with the results in Tables \ref{artifTab}, \ref{artifTab2} and \ref{artifTab3}. These results show that stationary dependence models for non-stationary spatial processes routinely provide a better fit if the deformation methodology is used as a preprocessing step. We also show that the original \cite{smith} deformation is outperformed by our extensions.

\subsection{Non-stationary Brown-Resnick and inverted Brown-Resnick process}
\label{NSMSPsec}
The first setting we consider consists of replications of a non-stationary Brown-Resnick, and inverted Brown-Resnick, process sampled at $64$ equally spaced locations on $[-1,1]\times[-1,1]$. We use a non-stationary variogram in the exponent function in \eqref{chiBR} to ensure that $\chi(h_{ij})$ is not simply a function of distance. In the context of non-stationary Gaussian processes, \citet{fouedjio2015estimation} propose a semivariogram of the form $\gamma^*(s_i,s_j)$ where
\begin{equation}
\label{nsRBF}
\gamma^*(s_i,s_j)=\gamma(\|\psi(s_j)-\psi(s_j)\|),
\end{equation}
and
\[
\psi(s)=o+(s-o)\|s-o\|
\]
is a radial basis function with some centre point $o$ and $\gamma(\cdot)$ is the stationary and isotropic semivariogram given in \eqref{vario}. The use of the radial basis function $\psi(s)$ within this semivariogram causes pairs that are closer to $o$ to be more strongly dependent than those pairs that are further away. From \eqref{chiBR} and \eqref{vario}, the Brown-Resnick process with this semivariogram has theoretical $\chi(s_i,s_j)$ given by
\begin{equation}
\label{chiNSMSP}
\chi(s_i,s_j)=2-2\Phi\left\{\frac{\|\psi(s_i)-\psi(s_j)\|^{\kappa/2}}{\lambda^{\kappa/2}\sqrt{2}}\right\},
\end{equation}
for locations $s_i, s_j$ and $\kappa \in (0,2],\lambda >0$. For this study, we take the centre $o$ to be the origin and use scale and shape parameters $\lambda=2$ and $\kappa=0.8$ in \eqref{chiNSMSP}. To illustrate the process a high resolution realisation is given in  \sfref{highres}. Simulations are produced using the method of \citet{Dieker2015}.\par
 \begin{table}[H]
\centering
   \begin{tabular}{c|c|l|r|c} 
 \hline
\shortstack{Process\\ (G-plane)}&\shortstack{Fitted Model\\ (D-plane)}&\shortstack{Deformation \\Method}& \shortstack{Proportion of \\lowest CLAIC}&\\
 \hline
 \multirow{5}{*}{\shortstack{Non-stationary\\ Brown-Resnick}}&\multirow{5}{*}{\shortstack{Stationary\\ Brown-Resnick}}&None&0&\\
 \cline{3-5}
 &&$\chi^{BR}$&0.22 &\multirow[|c|]{2}{*}{0.34}\\
& &$\chi^{IBR}_{q}$&0.12&\\
 \cline{3-5}
 &&$\rho$&0.44&\multirow{2}{*}{0.66}\\
& &\citet{smith} &0.22&\\
 \hline
 \multirow{5}{*}{\shortstack{Non-stationary\\ Inverted\\ Brown-Resnick}}&\multirow{5}{*}{\shortstack{Stationary\\ Inverted\\ Brown-Resnick}}&None&0\\
 \cline{3-5}
 &&$\chi^{BR}$&0.24 &\multirow{2}{*}{0.56}\\
& &$\chi^{IBR}_{q}$&0.32&\\
 \cline{3-5}
& &$\rho$&0.28&\multirow{2}{*}{0.44}\\
& &\citet{smith} &0.16&\\
 \hline
 \end{tabular}
 \caption{Proportion of lowest CLAIC estimates provided by fitting models to deformations for 50 realisations of non-stationary Brown-Resnick and inverted Brown-Resnick processes. The CLAIC has been estimated with a block size of $b=1$, corresponding to temporal independence. Composite likelihoods are estimated with the threshold in \eqref{compLikeq} as the $90\%$ empirical quantile, which is also used for estimating $\chi(h^{*}_{ij})$ in \eqref{chiestim}.}
 \label{artifTab}
\end{table}
Table \ref{artifTab} contains some interesting results. Most notably, in all cases a deformation has aided in model fitting when compared to using the original simulation grid. For both the max-stable, and inverted max-stable, cases, improvements on the efficacy of the original \citet{smith} method are made by utilising the Frobenius norm in the objective function. However, it is not entirely clear whether use of an extremal dependence measure for creating deformations is necessary in this case. We often found that deforming the space using measures for dependence throughout the distribution created better deformations than those using extremal dependence measures. We believe that this is because the variance of the estimator for $\rho(h^{*}_{ij})$ is much lower than that of $\chi(h^{*}_{ij})$, as we use all of the data to estimate correlation, and that there are strong similarities in patterns of spatial non-stationarity for the central- and extremal-dependence structures of this process. We next consider other processes with more complicated dependence structures.
\subsection{Max-mixture process}
\label{maxmixsec}
We now consider the hybrid dependence model, detailed in full by \cite{invertref}. Let $X(s)$ be a max-stable process and $Y(s)$ an asymptotically independent spatial process, both with standard Fr\'echet margins. For $\omega \in [0,1]$, $H(s)=\max\{\omega X(s), (1-\omega)Y(s)\}$
is an asymptotically dependent spatial process with standard Fr\'echet margins. In particular, we take $X(s)$ to be the non-stationary Brown-Resnick process detailed in Section \ref{NSMSPsec} and $Y(s)$ to be a marginally transformed stationary Gaussian process with the Mat\'ern correlation structure detailed in \eqref{matern}.\par
It can be shown that the theoretical $\chi(h_{ij})$ values for $H(s)$ are the same as for $X(s)$, but multiplied by $\omega$. There is no closed form for the correlation for $H(s)$ on the Gaussian scale. Computationally, it can be shown that it is a mixture of the correlation from both $X(s)$ and $Y(s)$. As such, we would expect the extremal dependence and central dependence of $H(s)$ to be mixtures of those coming from $X(s)$ and $Y(s)$, with different amounts of mixing occurring for both. We set $\omega$ to be $0.3$ and take $(\theta_1,\theta_2) = (1,1.2)$ in \eqref{matern}.\par
By construction of $H(s)$, taking its reciprocal creates an asymptotically independent process on standard exponential margins, as with the inverted max-stable process. As in Section \ref{NSMSPsec}, the simulation study is repeated separately for the asymptotically dependent and asymptotically independent mixtures. The results are given in Table \ref{artifTab2}.
\begin{table}[H]
\centering
   \begin{tabular}{c|c|l|r|c} 
 \hline
\shortstack{Process\\ (G-plane)}&\shortstack{Fitted Model\\ (D-plane)}&\shortstack{Deformation \\Method}& \shortstack{Proportion of \\lowest CLAIC}&\\
 \hline
\multirow{5}{*}{\shortstack{Asymptotically-dependent\\ Max-mixture}}&\multirow{5}{*}{\shortstack{Stationary\\ Brown-Resnick}}&None&0.06&\\
 \cline{3-5}
 &&$\chi^{BR}$&0.14 &\multirow[|c|]{2}{*}{0.78}\\
& &$\chi^{IBR}_{q}$&0.64&\\
 \cline{3-5}
& &$\rho$&0.16&\multirow{2}{*}{0.16}\\
& &\citet{smith} &0&\\
 \hline
\multirow{5}{*}{\shortstack{Asymptotically-independent\\ Max-mixture}}&\multirow{5}{*}{\shortstack{Stationary\\ Inverted\\ Brown-Resnick}}&None&0\\
 \cline{3-5}
& &$\chi^{BR}$&0.42 &\multirow{2}{*}{0.90}\\
 &&$\chi^{IBR}_{q}$&0.48&\\
 \cline{3-5}
 &&$\rho$&0.06&\multirow{2}{*}{0.10}\\
 &&\citet{smith} &0.04&\\
 \hline
 \end{tabular}
 \caption{Proportion of lowest CLAIC estimates provided by fitting models to deformations of 50 realisations of asymptotically dependent and asymptotically independent max-mixture processes. The CLAIC has been estimated with a block size of $b=1$, corresponding to temporal independence. Composite likelihoods are estimated with the threshold in \eqref{compLikeq} as the $90\%$ empirical quantile, which is also used for estimating $\chi(h^{*}_{ij})$ in \eqref{chiestim}.}
 \label{artifTab2}
\end{table}
In contrast to the results given in Table \ref{artifTab}, Table \ref{artifTab2} shows a clearer need for an extremal dependence-based approach when creating deformations for a process that exhibits more complicated dependence structures. Here this max-mixture process is designed to represent a process with a mixture of 
stationarity in both the extremal dependence and dependence throughout the distribution. We now consider a process that has non-stationary extremal dependence, but is nearly stationary in the body.
\subsection{Gaussian mixture process}
\label{gaussmixproc}
With previous simulations, we found it is sometimes sufficient to simply use measures of central dependence when deforming the spatial domain to create a process with a more stationary extremal dependence structure. This is because the central- and extremal-dependence structures of these processes are closely related and using either approach typically creates similar deformations. In applications, we may find that these structures are not so closely related. As such, we are motivated to consider a process that is designed to have completely different dependence in the body  to the tails.\par
Let $Y_{\textsc{S}}(s),Y_{\textsc{NS}}(s)$ be stationary and non-stationary Gaussian processes, respectively, each with standard Gaussian margins. We then consider the process
\begin{equation}
\label{gaussmixeq}
Y^*(s)=\begin{cases}
Y_{\textsc{S}}(s),&\text{if }\Phi(Y(s_0))\leq p\\
Y_{\textsc{NS}}(s),&\text{if }\Phi(Y(s_0))> p
\end{cases},
\end{equation}
where $s_0 \in S$ is a fixed location, $\Phi(\cdot)$ is the standard Gaussian cdf, and $p \in [0,1]$ is a probability. By specifying $Y^*(s)$ in this manner, we create a process with an extremal dependence structure determined mostly by the correlation structure of $Y_{\textsc{NS}}$ and with dependence through the body determined mostly by $Y_{\textsc{S}}$. Simulation of this process is simple; we draw $Y(s_0) \sim N(0,1)$ and then simulate the rest of the field conditioning on that value and whether $\Phi(Y(s_0)) \leq p$ or $\Phi(Y(s_0)) > p$. \par
For this particular study, we use replications of this Gaussian mixture sampled at $81$ equally spaced locations in $[-1, 1] \times [-1,1]$. We take $s_0$ to be the origin and $p=0.9$. Both $Y_{\textsc{S}}$ and $Y_{\textsc{NS}}$ are specified to have the Mat\'ern correlation structure given in \eqref{matern}, with respective parameter sets $\boldsymbol{\theta}^{({\textsc{S}})}=(\theta^{({\textsc{S}})}_1,\theta^{({\textsc{S}})}_2)$ and $\boldsymbol{\theta}^{(\textsc{NS})}=(\theta^{(\textsc{NS})}_1,\theta^{(\textsc{NS})}_2, o )$. Note that $\boldsymbol{\theta}^{(\textsc{NS})}$ contains an extra parameter as we use the difference of the radial basis functions given in \eqref{nsRBF} and detailed by \citep{fouedjio2015estimation} as a measure of pairwise distance, rather than Euclidean distance. The parameters for this study are set to $\boldsymbol{\theta}^{({\textsc{S}})}=(2,1)$ and $\boldsymbol{\theta}^{(\textsc{NS})}=(2,0.8,(0,0))$. Results are given in Table \ref{artifTab3}.
 
 \begin{table}[H]
\centering
   \begin{tabular}{c|c|l|r|c} 
 \hline
\shortstack{Process\\ (G-plane)}&\shortstack{Fitted Model\\ (D-plane)}&\shortstack{Deformation \\Method}& \shortstack{Proportion of \\lowest CLAIC}&\\
 \hline
 \multirow{5}{*}{\shortstack{Gaussian\\ Mixture}}&\multirow{5}{*}{\shortstack{Stationary\\ Inverted\\ Brown-Resnick}}&None&0&\\
  \cline{3-5}
& &$\chi^{BR}$& 0.08&\multirow[|c|]{2}{*}{1}\\
& &$\chi^{IBR}_{q}$&0.92&\\
  \cline{3-5}
& &$\rho$&0&\multirow{2}{*}{0}\\
& &\citet{smith} &0&\\
 \hline
 \end{tabular}
 \caption{Proportion of lowest CLAIC estimates provided by fitting models to deformations of 50 realisations of the Gaussian mixture process, see  \eqref{gaussmixeq}. The CLAIC has been estimated with a block size of $b=1$, corresponding to temporal independence. Composite likelihoods are estimated with the threshold in \eqref{compLikeq} as the $90\%$ empirical quantile, which is also used for estimating $\chi(h^{*}_{ij})$ in \eqref{chiestim}.}
 \label{artifTab3}
\end{table}
Table \ref{artifTab3} highlights a clear need for extremal dependence-based methods when creating deformations for processes that have different patterns of non-stationarity in their central- and extremal dependence structures. In contrast to the results given in the previous studies, here using $\chi(h^{*}_{ij})$ or $\chi_{q}(h^{*}_{ij})$ is always favoured.
  \section{Case studies}
  \label{casestud}
  We present two case studies using our deformation methodology. In both cases, we follow the procedure set out in Section \ref{secpract}. However, as we consider relatively large spatial domains we use Great Earth distance in place of Euclidean distance for $h$ and $h^*$. We consider $30$ different initial index sets, taking the best deformation over all sets. Here we define the best deformation to be that which provides the lowest objective value in \eqref{frob} whilst remaining a bijective mapping. When using extremal dependence measures, we focus on deformations based on $\chi^{BR}$ only, following the justification in Section \ref{secmodelspec}. We then fit max-stable and inverted max-stable models to the data in the G-plane and D-plane, comparing the model fits using CLAIC estimates. For both studies, all pairs of sampling locations are used in model fitting and the block size in \eqref{scoreest} corresponds to a season. We propose two diagnostics for scrutinising the model fits and deformations.
  \subsection{Australian summer temperatures}
  \label{secheat}
\begin{figure}[H]
 
  \centering
  
 \includegraphics[width=\linewidth]{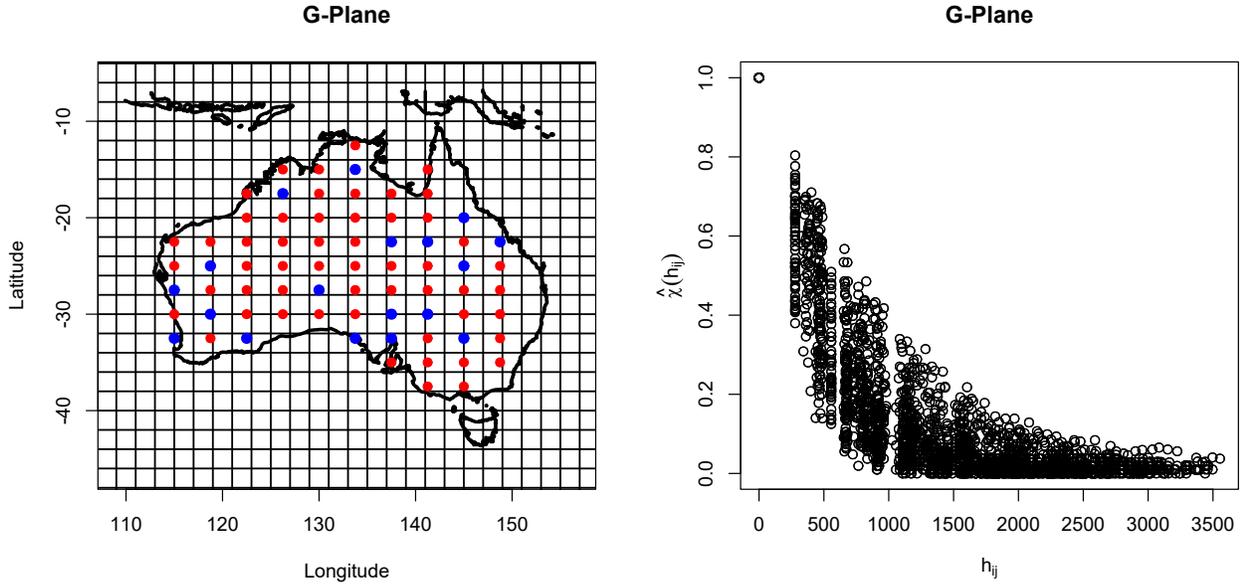} 

  \caption{Australia summer temperatures. Left: the original 72 sampling locations. The blue points are the anchor points used for the thin-plate splines. Right: empirical $\chi(h_{ij})$ measures against distance (km). Estimates $\hat{\chi}(h_{ij})$ are calculated above a threshold given by the $98\%$ empirical quantile.}
   \label{heatG}
  \end{figure}

\begin{figure}[H]
  
  \centering
  
\includegraphics[width=\linewidth]{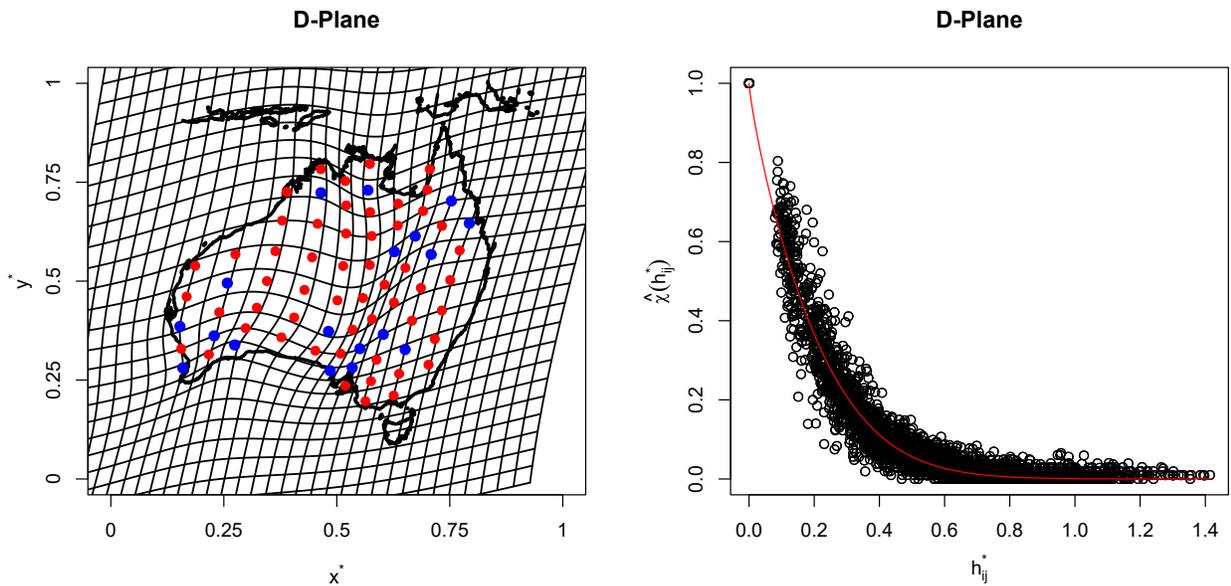} 
 
    \caption{Australia summer temperatures. Left: the 72 sampling locations in the D-plane. The blue points are the anchor points used for the thin-plate splines. The coordinates have been scaled to $[0,1]\times[0,1]$, which equals the aspect ratio of the left plot in Figure \ref{heatG}. Right: empirical $\chi(h^{*}_{ij})$ measures against distance in the D-plane. The red line gives the fitted function $\chi(h^{*})$ used in the deformation.}
 \label{heatD}
  \end{figure}

Data consist of daily summer (DJF) maximum near-surface air temperatures taken from the HadGHCND global gridded dataset \citep{doi:10.1029/2005JD006280} and interpolated to 72 grid point locations covering Australia, for the period 1957-2014. Previous analysis of this data has been conducted using the multivariate conditional extremes model, detailed in Section \ref{ConExSec} \citep{hugo} and its spatial extension \citep{wadsworth2018spatial}.
 Figure \ref{heatG} shows the original sampling locations and estimated pairwise $\chi(h_{ij})$ against distances. We estimate $\chi(h_{ij})$ by setting $q=0.98$ in \eqref{chiestim}. The deformation was produced using $m^*=18$, i.e.\ a quarter of the original sampling locations. These are presented as the blue points on Figures $\ref{heatG}$ and $\ref{heatD}$, where the latter figure depicts the sampling locations in the D-plane. Figure $\ref{heatD}$ also presents $\hat{\chi}(h^{*}_{ij})$ against distance in the deformed space. We observe that the deformation has created a process that appears to be much more stationary with regards to the $\chi(h^{*}_{ij})$ estimates in the new coordinates.\par
   
\begin{table}[H]
\centering
   \begin{tabular}{l|l|c|c|c} 
 \hline
&Model & Negative Composite Log-Likelihood $(\times 10^6)$&$(\hat{\kappa},\hat{\lambda})$ (2 d.p.)&CLAIC $(\times 10^7)$\\
 \hline
 \multirow{2}{*}{G-Plane}&IMSP$^*$&3.078&(2.00, 1048.20)&$6.157$\\
  &MSP&3.078&(1.59, 358.30)&6.157\\
 \hline
  \multirow{2}{*}{D-Plane}&IMSP$^*$&3.074&(2.00, 2.61)&$6.148$\\
 &\textbf{MSP}&3.073&(1.71, 0.95)&\textbf{6.146}\\
 \hline
 \end{tabular}
 \caption{Model parameters and diagnostics for the Australian summer temperatures data. Composite likelihoods are estimated with the threshold in \eqref{compLikeq} as the $98\%$ empirical quantile. ($^*$ estimated using Smith process likelihood). CLAIC and negative composite log-likelihood estimates are given to four significant figures. }
 \label{heatTab}
\end{table}
The fits of the max-stable and inverted max-stable models are summarised in Table \ref{heatTab}.
The CLAIC estimates suggest that a max-stable model is more appropriate for the data. This becomes even more apparent when we consider that fitting an inverted Brown-Resnick model yields an inverted Smith model as the best fit. These processes are typically quite smooth and often provide unrealistic representations of actual data. However, we note that when naively fitting models on the G-plane, the inverted Smith model provided the lowest CLAIC estimate. This is further evidence that non-stationarity in this data should be incorporated into the modelling procedure.\par
We use two diagnostics to scrutinise the deformation and the model fit. As our deformation method is tailored to $\chi(h^{*}_{ij})$, we seek to use other extremal dependence measures to verify that the resulting deformation is not subject to overfitting. To do this,
the conditional extremes model described in Section \ref{ConExSec} is fitted pairwise and the parameter estimates are used to calculate the conditional expectation of one variable when the other variable is at the modelling threshold $u$, taken as the 98\% quantile of the marginal distribution. For each pair, $(X(s_i),X(s_j)),i \neq j$, we have
\[
\mathbb{E}\left[X(s_j)|X(s_i)=u\right]=\hat{\alpha}u+u^{\hat{\beta}}\hat{\mu},
\]
where $(\hat{\alpha},\hat{\beta},\hat{\mu})$ are the maximum likelihood estimates for the model. For a stationary and isotropic process, we would expect this measure to be a smooth function of Euclidean distance. The conditional expectation is plotted against distance for both the process on the G-plane and the D-plane.\par
A second diagnostic is used to evaluate the best model fit in the D-plane. As we have used $\chi(h^{*}_{ij})$ to create the deformations, we compare the theoretical triple-wise $\chi$, which we denote $\chi(s^*_i,s^*_j,s^*_k)=\chi(s_i,s_j,s_k)$, from the model fits against empirical estimates. The triple-wise $\chi$ is defined as
\begin{align*}
\chi(s^*_i,s^*_j,s^*_k)&=\lim_{q\rightarrow 1}\Pr[F_{i}\{Z(s^*_i)\}>q,F_{j}\{Z(s^*_j)\}>q|F_{k}\{Z(s^*_k)\}>q]\\
&=\lim_{q\rightarrow 1}\Pr[F_{i}\{Z(s_i)\}>q,F_{j}\{Z(s_j)\}>q|F_{k}\{Z(s_k)\}>q]=\chi(s_i,s_j,s_k)
\end{align*}
for $i \neq j \neq k$. For a Brown-Resnick process, the theoretical value for this measure is
\[
\chi(s^*_i,s^*_j,s^*_k)=3-V_2(1,1;i,j)-V_2(1,1;i,k)-V_2(1,1;j,k)+V_3(1,1,1),
\]
where $V_2(\cdot,\cdot;l,m)$ is the pairwise exponent given in \eqref{exponent} and $V_3(\cdot,\cdot,\cdot)$ is the triple-wise exponent measure, for which the parametric form is given in \citet{brpro}; recall that if the process is stationary, both of these are functions of Euclidean distance. A similar parametrisation can be given for $\chi_q(s^*_i,s^*_j,s^*_k)$ for an inverted Brown-Resnick process, which is $\chi_q(s^*_i,s^*_j,s^*_k)=(1-q)^{V_3(1,1,1)-1}$.   \par

Standard errors for empirical estimates of $\chi_q(s^*_i,s^*_j,s^*_k)$ are estimated using a stationary bootstrap \citep{statboot}. We begin by drawing a random block size $B$ from a geometric distribution with mean $K$. The bootstrap sample for locations $s_i,s_j,s_k,\; i \neq j \neq k$ is built by drawing a random starting time $\tau$ and creating a block of observations
\[
\{\mathbf{z}^{*}_{\tau},\dots,\mathbf{z}^{*}_{\tau+B-1}\},\;\;\;\text{where}\;\;\;\mathbf{z}^{*}_t=\{z_t(s_i),z_t(s_j),z_t(s_k)\},
\]  
which we add to the bootstrap sample. This procedure is repeated and the bootstrap is built up iteratively until it has length $n$. We then estimate $\chi(s^*_i,s^*_j,s^*_k)$ for that sample and repeat for a number of samples.
 When choosing locations to compare empirical and theoretical values of $\chi(s^*_i,s^*_j,s^*_k)$, we take advantage of the gridded structure of the coordinates in the G-plane, and ensure that each set of points share roughly the same configuration and pairwise distances. This is used to evaluate the stationarity of the dependence structure on the original G-plane, as we would expect the empirical values of $\chi(s^*_i,s^*_j,s^*_k)$ to be consistently similar across sets of locations with the same configuration.
\par
\begin{figure}[H]
 
  \centering

   \includegraphics[width=\linewidth]{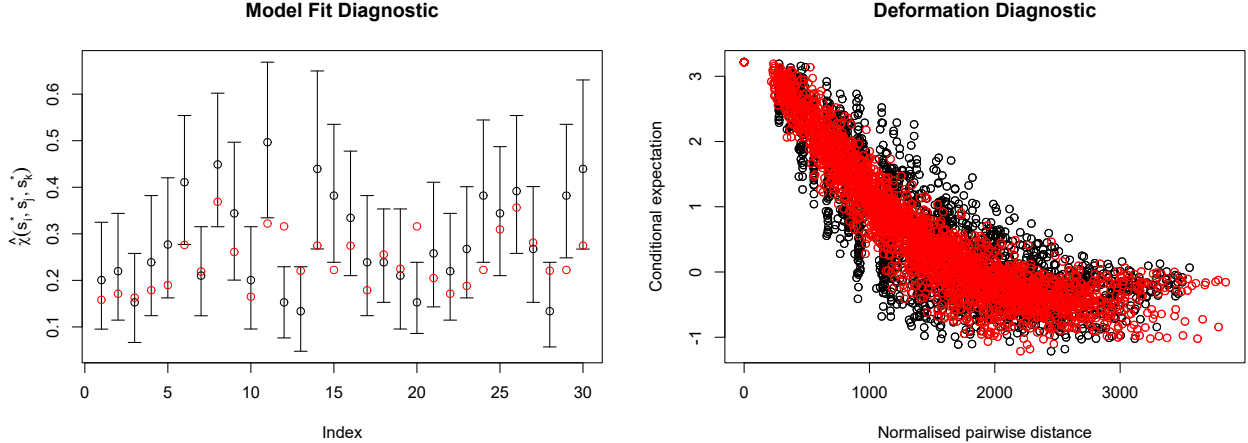} 
  
  \caption{Australian summer temperatures diagnostics. Left: estimates of $\chi(s^*_i,s^*_j,s^*_k)$ (black dots) and $95\%$ confidence intervals using the stationary bootstrap. Red dots are the respective theoretical values suggested by the model fit. Right: conditional expectation from conditional extremes model. Red points denote estimates for the process on the D-plane; black points are those on the G-plane.}
   \label{heatDiag}
  \end{figure}

Diagnostics for the deformations and best model fit are given in Figure \ref{heatDiag}. For the estimation of $\chi(s^*_i,s^*_j,s^*_k)$, 30 sets of three adjacent locations along the north/south transect in the G-plane are randomly selected and a stationary bootstrap with mean block size $K=14$ and $1000$ samples is used to create $95\%$ confidence intervals for the empirical estimates of $\chi(s^*_i,s^*_j,s^*_k)$.  Empirical estimates of $\chi(s^*_i,s^*_j,s^*_k)$ are calculated above the $98\%$ quantile. The right panel of Figure \ref{heatDiag} displays estimates for the conditional expectation from the conditional extremes model, where distances are normalised so that the average distance is equal for both the values in the G-plane and the D-plane.
  The diagnostic based on $\chi(s^*_i,s^*_j,s^*_k)$ from Figure \ref{heatDiag} suggests that a max-stable model is a reasonable fit for the data in the deformed space, as the patterns of the theoretical $\chi(s^*_i,s^*_j,s^*_k)$ values follow the empirical estimates. The large variability in the bootstrap estimates across sets of locations with similar configurations suggests that the process on the original plane is highly non-stationary. Estimates from the conditional extremes model provide further evidence that the deformation has produced something more stationary with regards to the dependence structure, especially at smaller distances. The use of a measure for extremal dependence that is not used for fitting lends credibility to the $\chi(h^*_{ij})$ plot in Figure \ref{heatD} and suggests that the deformation has worked well.
  \subsection{UK precipitation rate}
  \label{secrain}

Data consist of hourly precipitation rate (mm/day) observed at locations on two $10 \times 10$ grids; the first is centred in Snowdonia, Wales and the second is centred in the Scottish Highlands. Observations are taken from the UK climate projections 2018 (UKCP18) \citep{lowe2018ukcp18} which contain values produced at hourly intervals on $2.2 \times 2.2$km$^2$ grid boxes between the years 1980 and 2000. We have treated the centre of each grid box as a sampling location and we take every fifth grid box to create the $10 \times 10$ grid of sampling locations. Observations are aggregated to 12-hr intervals, beginning at 12pm, and to remove the seasonal effect often observed in precipitation data, we have taken only winter observations (DJF). This leaves 3600 observations at each sampling location.
\par
\begin{figure}[H]
 
  \centering
  
 \includegraphics[width=\linewidth]{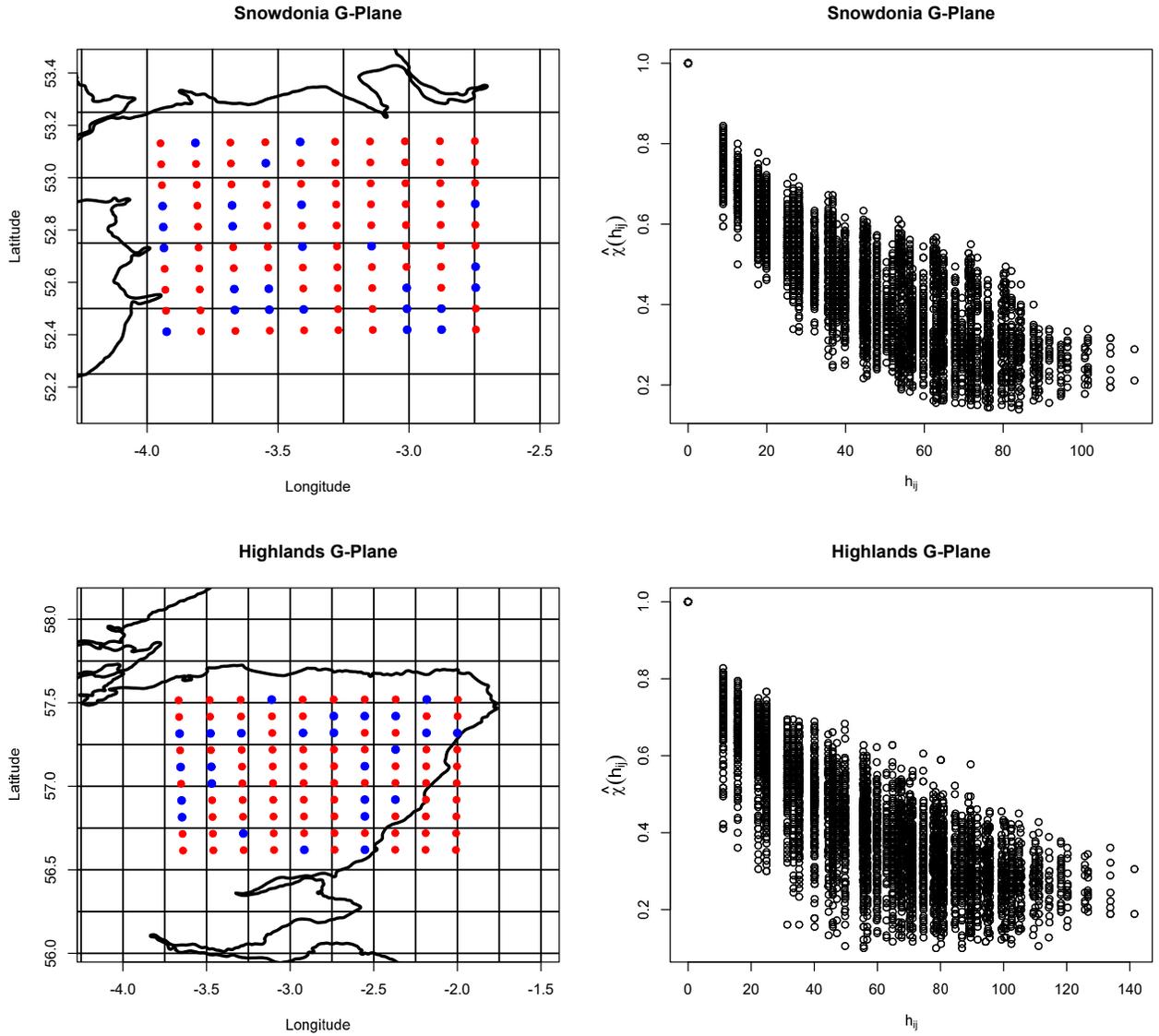} 

  \caption{Top row: Snowdonia. Bottom row: Scottish Highlands. Left: the original 100 sampling locations. The blue points are the anchor points used for the thin-plate splines. Right: empirical $\chi(h_{ij})$ measures against distance (km) in the respective G-planes. Estimates $\hat{\chi}(h_{ij})$ are calculated above a threshold given by the $95\%$ empirical quantile.}
   \label{rainG}
  \end{figure}
\begin{figure}[H]
 
  \centering
  
 \includegraphics[width=\linewidth]{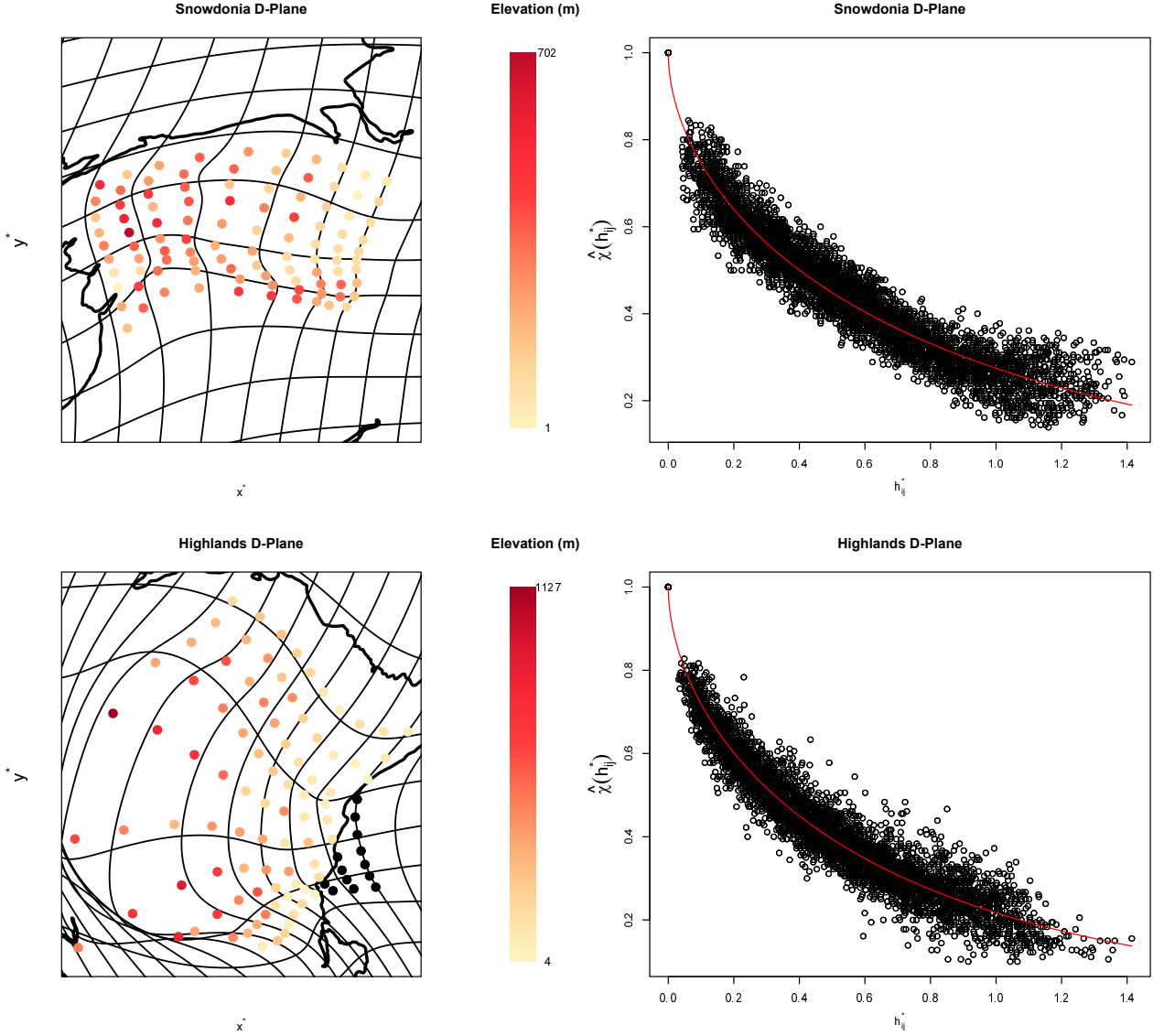} 

  \caption{Top row: Snowdonia. Bottom row: Scottish Highlands. Left: the 100 sampling locations in their respective D-planes. The points are coloured such that darker points correspond to sampling locations with higher elevation and black points correspond to locations over sea. The coordinates have been scaled to $[0,1]\times[0,1]$, which equals the aspect ratio of the left plots in Figure \ref{rainG}. Right: empirical $\chi(h^{*}_{ij})$ measures against distance in the D-plane. The red line gives the fitted function $\chi(h^{*})$ used in the deformation.}
   \label{rainD}
  \end{figure}

 Figure \ref{rainG} shows both sets of original sampling locations and their respective estimates of $\chi(h_{ij})$ against distances. In both cases, we estimate $\chi(h_{ij})$ by setting $q=0.95$ in \eqref{chiestim}. Both deformations are produced using $m^*=25$ and these are presented as the blue points in Figure $\ref{rainG}$. Figure $\ref{rainD}$ presents both deformations and estimates of $\hat{\chi}(h^{*}_{ij})$ against distance in the respective deformed spaces. We observe that both deformations have created a process that appears to be much more stationary with regards to their respective $\chi(h^{*}_{ij})$ estimates in the new coordinates. In both cases, deformations are more prominent around areas of higher elevation.  \par

\begin{figure}[H]
 
  \centering

   \includegraphics[width=\linewidth]{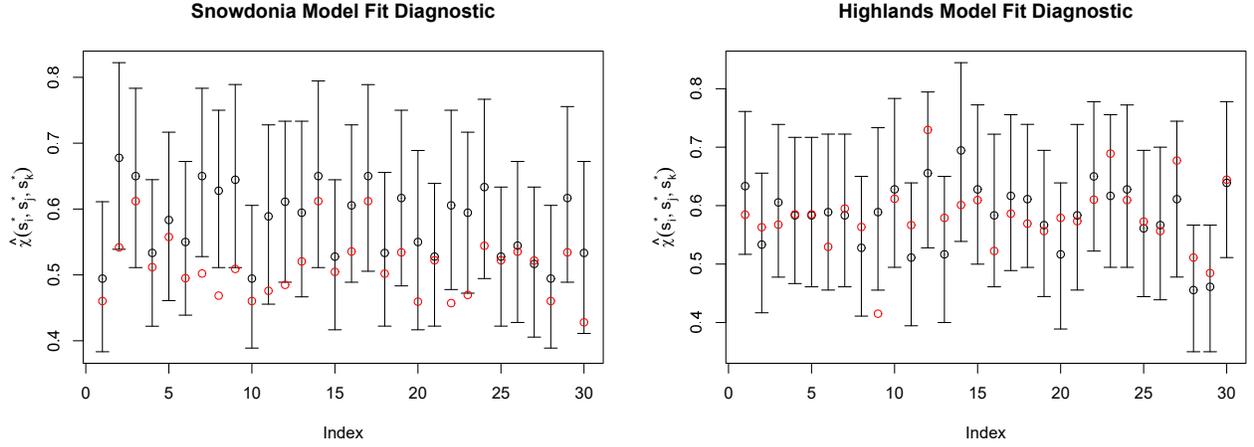} 
  \caption{UK precipitation model fit diagnostics. Estimates of $\chi_{q}(s^*_i,s^*_j,s^*_k)$ (black dots) with $q=0.95$ and $95\%$ confidence intervals using the stationary bootstrap. Red dots are the respective theoretical values suggested by the model fits. Left: Snowdonia. Right: Highlands. }  
   \label{rainDiagStatBoot}
  \end{figure}

\begin{figure}[H]
 
  \centering

   \includegraphics[width=\linewidth]{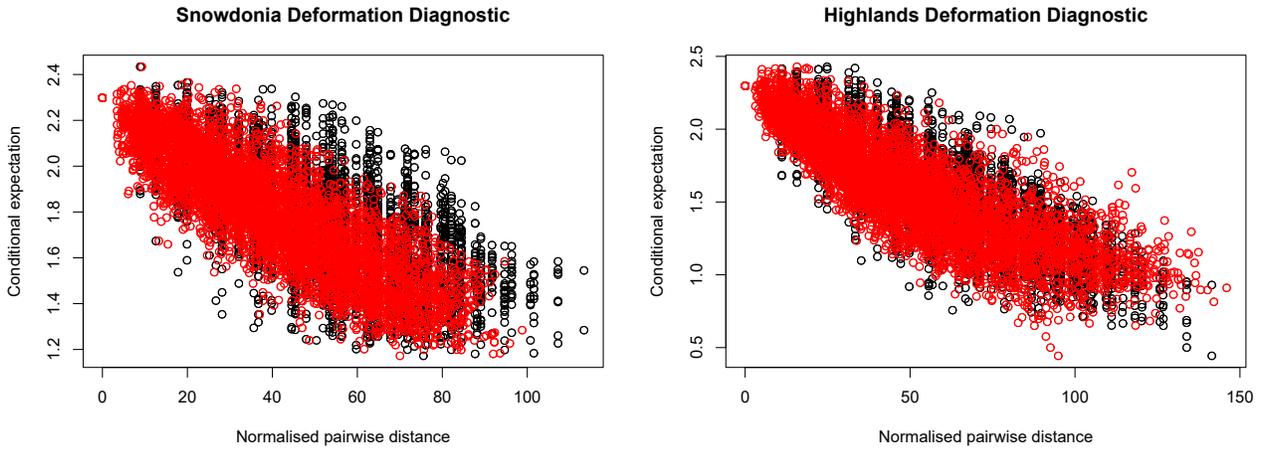} 
  \caption{UK precipitation deformation diagnostics. Conditional expectation from conditional extremes model. Red points denote estimates for the process on the D-plane; black points are those on the G-plane. Left: Snowdonia. Right: Highlands. }  
   \label{rainDiagCondExp}
  \end{figure}

 \begin{table}[H]
\centering
   \begin{tabular}{l|c|c|c|c|c} 
 \hline
&&\multirow{2}{*}{Model} & Negative Composite&\multirow{2}{*}{$(\hat{\kappa},\hat{\lambda})$ (2 d.p.)} &\multirow{2}{*}{CLAIC ($\times 10^7$)}\\
&& &  Log-Likelihood ($\times 10^6$)&&\\
 \hline
 \multirow{4}{*}{Snowdonia}&\multirow{2}{*}{G-Plane}&IMSP&8.023&(1.40, 111.84)&1.605\\
 &&MSP&8.050&(1.00, 25.96)&1.610\\
\cline{2-6}
  &\multirow{2}{*}{D-Plane}&\textbf{IMSP}&8.011&(1.29, 3.33)&\textbf{1.602}\\
 &&MSP&8.037&(0.93, 0.69)&1.607\\
 \hline
 \multirow{4}{*}{Highlands}&\multirow{2}{*}{G-Plane}&IMSP&8.099&(1.25, 143.77)&1.620\\
 &&MSP&8.124&(0.87, 27.37)&1.625\\
\cline{2-6}
  &\multirow{2}{*}{D-Plane}&\textbf{IMSP}&8.076&(1.30, 3.34)&\textbf{1.615}\\
 &&MSP&8.099&(0.93, 0.69)&1.620\\
 \hline
 \end{tabular}
 \caption{Model parameters and diagnostics for the UK precipitation data. Composite likelihoods are estimated with the threshold in \eqref{compLikeq} as the $95\%$ empirical quantile. CLAIC and negative composite log-likelihood estimates are given to four significant figures. }
 \label{rainTab}
\end{table}
Table \ref{rainTab} summarises the fits for the Brown-Resnick and inverted Brown-Resnick models for both sets of sampling locations. The CLAIC estimates in Table \ref{rainTab} suggest that an inverted max-stable model is the most appropriate for both the Snowdonia and Highlands data. Both see improved fits using the sampling locations mapped to the respective D-planes. In Figures \ref{rainDiagStatBoot} and \ref{rainDiagCondExp}, we present diagnostics for the deformations and best model fits using the same measures described in Section \ref{secheat}. As the best fitting model for both datasets is the inverted Brown-Resnick process, Figure \ref{rainDiagStatBoot} compares empirical estimates and model-based values of $\chi_{q}(s^*_i,s^*_j,s^*_k)$ with $q=0.95$. Confidence intervals for the empirical estimates of $\chi_{q}(s^*_i,s^*_j,s^*_k)$ are calculated by randomly selecting 30 sets of three adjacent locations along the east/west transect and a using stationary bootstrap with mean block size $K=14$ and $1000$ samples.  For the diagnostic given in Figure \ref{rainDiagCondExp}, the $95\%$ quantile is used for fitting the conditional extremes model and we plot the pairwise conditional expectation estimates against distance. Distances are normalised so that the average distance is equal for both the values in the G-plane and the D-plane.

  
Figure \ref{rainDiagStatBoot} shows that the inverted max-stable model gives a relatively good fit to the extremal dependence of both datasets with sampling locations mapped to their respective D-planes, but the fit appears better for the Scottish Highlands. The low variability in the $\chi_{q}(s^*_i,s^*_j,s^*_k)$ estimates suggests that the original process may not be highly non-stationary. 
The pairwise conditional expectation estimates in Figure \ref{rainDiagCondExp} suggest that both deformations have produced a more stationary process, albeit more so in the case of the Snowdonia D-plane. The small change in the Highlands estimates may suggest that overfitting to the $\chi(h_{ij})$ values has occurred, especially when compared to the Snowdonia estimates. This may also explain the stronger agreement of the $\chi_{q}(s^*_i,s^*_j,s^*_k)$ measures in Figure \ref{rainDiagCondExp}. To investigate the possibility of overfitting, we recreated the diagnostic using deformations created with fewer anchor points, but this did not show any improvements.

  \section{Discussion}
  \label{discuss}
  In this paper, we presented a simple yet effective approach to modelling non-stationary extremal dependence. This approach extends that of \cite{sampandgut} and \cite{smith} to be applicable for modelling extremal dependence, rather than dependence throughout the body. We do this by replacing the objective function in these methods with the Frobenius norm of the difference between empirical, and theoretical, pairwise dependency matrices, with the theoretical measures coming from a stationary dependence model. Although most of our focus is on $\chi(h^{*}_{ij})$ as the dependence measure, we have also shown that this is easily replaced by other measures, such as $\chi_q(h^{*}_{ij})$ and correlation. Model selection is carried out using pairwise composite likelihoods and CLAIC estimation and we propose diagnostics for evaluating these model fits.\par
 We presented two case studies; in each scenario, we showed that when modelling the extremal dependence of the data using stationary models, better fits are provided using our methodology. Here we have fit very simple models to the data. However, in practice these deformations may be used as a pre-processing step to reveal covariates or orography that can be incorporated into the modelling procedure. Two diagnostics were introduced and used to provide evidence that our approach has produced a process which is more stationary with regards to the extremal dependence. \par 
 
 As with many areas of extreme value analysis, there is a bias-variance trade-off present when estimating $\chi_q$. Using values of $q$ closer to 1 puts greater focus on extremal dependence at the expense of increased variance of the estimator. In Sections \ref{simstudy} and \ref{casestud}, we choose $q$ close to 1 whilst preserving some initial spatial structure observed in the $\chi$ estimates. However, if $q$ is too high then it is possible that any structure is masked by the high variability of the estimators and the deformation methodology is likely to fail in such circumstances. We have not considered the effect of estimator variability on the deformation, but note this could form a future research direction.\par
 A further issue that could be considered is the possible non-bijectivity of the mapping used in the deformation. We detail an approach to reduce this in Section \ref{simstudy}, however, this method is not particularly robust. Bijectivity of deformations must be checked by eye which can become cumbersome when a large number are produced. To avoid this necessary supervision, the G-plane can be represented as a Delaunay triangulation, see \cite{doi:10.1198/1061860043100} and \cite{youngman2020flexible}. Incorporating this extra computational aspect into the model adds to the complexity, and so as to preserve the simplicity of our approach, we leave this as a future consideration.
\renewcommand{\abstractname}{Acknowledgements}
\begin{abstract}
We thank the associate editor and two referees for comments that improved the manuscript. We gratefully acknowledge funding through the STOR-i Doctoral Training Centre and Engineering and Physical Sciences Research Council (grant EP/L015692/1 and fellowship EP/P002838/1). The authors are grateful to Simon Brown and Robert Shooter of the Met Office Hadley Centre, UK, for access to data and preliminary code.
 \end{abstract}
 \renewcommand{\abstractname}{Data Accessibility}
\begin{abstract}
The data that supports the findings in this study are available in the accompanying \texttt{R} package, \texttt{sdfEXTREME}, which can be found at \url{https://github.com/Jbrich95/sdfExtreme}.
 \end{abstract}
 \bibliography{ref}
\section{Supplementary Material}
 \subsection{Comparison of \texorpdfstring{$\chi^{GG}(h)$}{x(h)} and  \texorpdfstring{$\chi^{BR}(h)$}{x(h)}}
\begin{figure}[h]
 \centering
  \includegraphics[width=\linewidth]{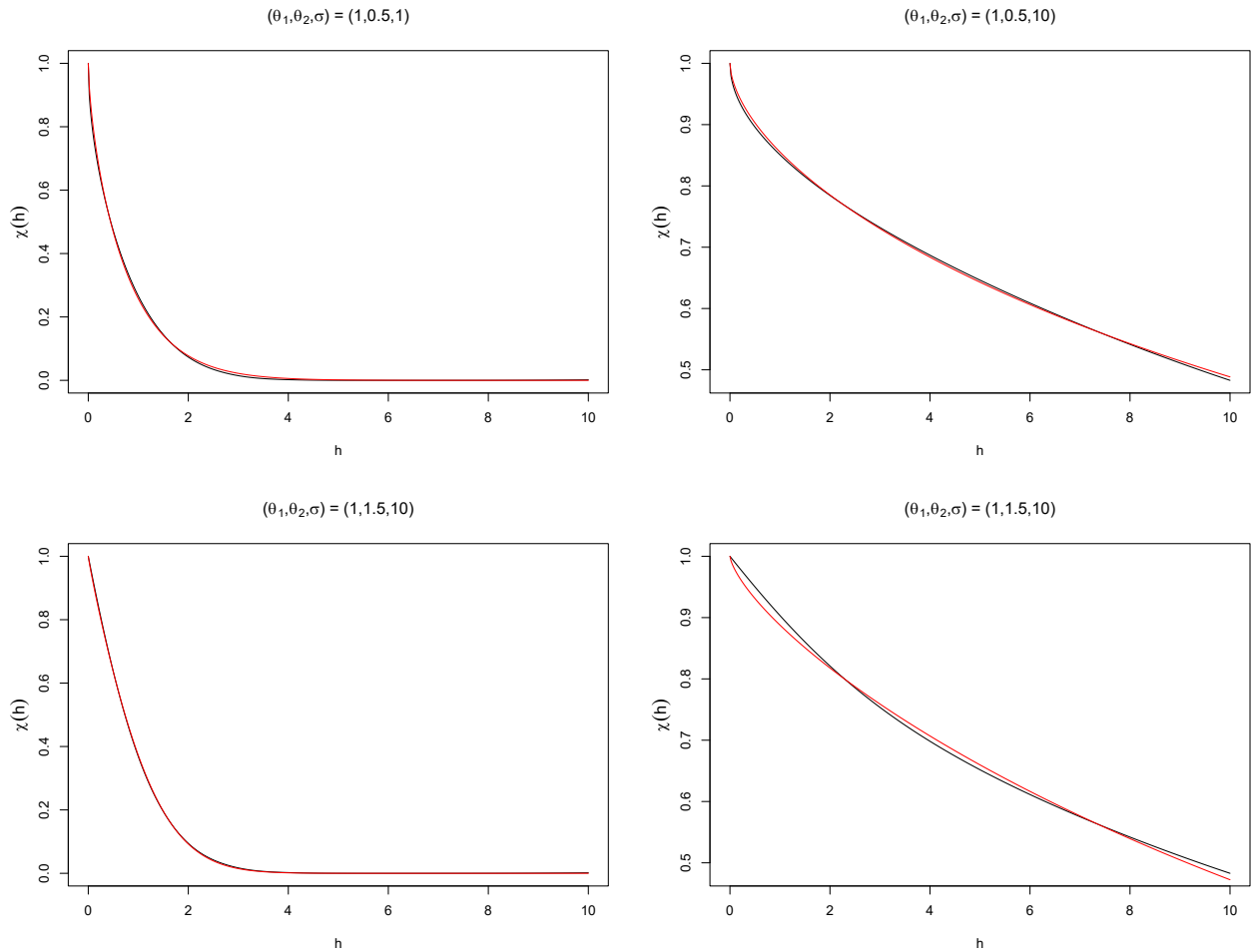} 
  \caption{Comparison of $\chi^{GG}(h)$ (black) and $\chi^{BR}(h)$ (red) for different parameter values.}
   \label{GG}
  \end{figure}

  In Section \ref{secmodelspec}, we discuss using the theoretical $\chi(h)$ function from a Brown-Resnick model rather than a Gaussian-Gaussian model. This is because the former is less computationally expensive to compute and often approximates the latter very closely for $h \in \mathbb{R}_{+}$. To illustrate this, Figure \ref{GG} shows the best approximation of $\chi^{BR}(h)$ to some fixed $\chi^{GG}(h)$ with Mat\'ern correlation function and parameter set $(\theta_1=1,\theta_2,\sigma)$. Here $\theta_1$ is set to $1$ as this controls spatial scaling only. The functions $\chi^{BR}(h)$ are produced by minimising $\|\chi^{BR}(h)-\chi^{GG}(h)\|_{F}$ for a sequence of fixed $h \in [0,10]$. Each figure uses different values of $(\theta_2,\sigma)$.
    \subsection{High resolution heatmap of non-stationary max stable process in Sections \ref{NSMSPsec} and \ref{maxmixsec}}
\begin{figure}[h]
\centering
\includegraphics[width=0.4\linewidth]{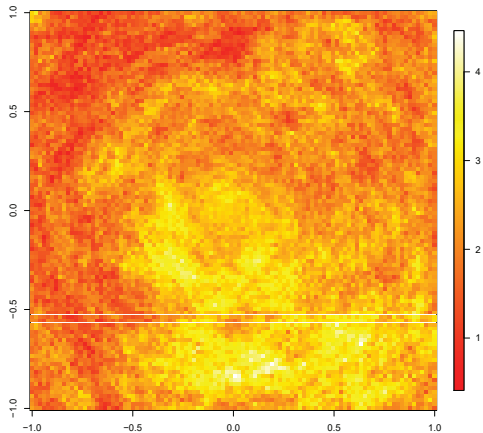} 
\caption{Heatmap of one high-resolution realisation of the max-stable process proposed in Section \ref{NSMSPsec} on a Gumbel marginal scale. This is sampled at $100 \times 100$ equally spaced points in $[-1,1]\times[-1,1]$. The parameter values in \eqref{chiNSMSP} are taken to be $o=(0,0)$ and we have $(\lambda,\kappa)=(2,0.8)$.}
\label{highres}
\end{figure}

Figure \ref{highres} gives a high-resolution heatmap of one realisation of a non-stationary Brown-Resnick process, with pairwise $\chi(s_i,s_j)$ given in \eqref{chiNSMSP}. This process is used in the simulation studies in Sections \ref{NSMSPsec} and \ref{maxmixsec}.
\end{document}